Type of Manuscript:
Research

# Comparison of forecasting of the risk of coronavirus (COVID-19) in high-quality and low-quality healthcare systems, using ANN models

**Aseel Sameer Mohamed[1], Nooriya A. Mohammed[2]**

[1] Family and Community Medicine Department, Al Kindy Medical College, University of Baghdad, aseelsameer@kmc.uobaghdad.edu.iq

[2] Planning and Studies Office, Ministry of Electricity, Baghdad, IRAQ, nooria.abed@gmail.com

## ABSTRACT

COVID-19 is a disease that has abnormal over 170 nations worldwide. The number of infected people (either sick or dead) has been growing at a worrying ratio in virtually all the affected countries. Forecasting procedures can be instructed so helping in scheming well plans and in captivating creative conclusions. These procedures measure the conditions of the previous thus allowing well forecasts around the state to arise in the future. These predictions strength helps to make contradiction of likely pressures and significances. Forecasting procedures production a very main character in elastic precise predictions. In this case study used two models in order to diagnose optimal approach by compared the outputs.

This study was introduced forecasting procedures into Artificial Neural Network models compared with regression model. Data collected from Al –Kindy Teaching Hospital from the period of 28/5/2019 to 28/7/2019 show an energetic part in forecasting. Forecasting of a disease can be done founded on several parameters such as the age, gender, number of daily infections, number of patient with other disease and number of death . Though, forecasting procedures arise with their private data of tests. This study chats these tests and also offers a set of commendations for the persons who are presently hostile the global COVID-19 disease.

## INTRODUCTION:

The pandemic of COVID-19 (or SARS-CoV-2), also recognized as the coronavirus pandemic, is presently causing chaos in more than 200 countries over the world. In February 2021, the number of the infected people with this virus reaches 113 million worldwide, the number of deaths reached 1.5 M by this infectious disease. The number infected people are growing every day [1]. Different nation's compulsory different methods to constrain the spread of COVID-19 including: mask use, isolation, limiting social meetings, closing of markets, frequently washing hands, cleaning frequently used areas, and travel limitations. Some countries use compulsory lockdowns to limit the eruption of COVID-19 [2-4].

This infectious extremely transmissible disease attacks the breathing system of the body. It is dispersal finished the physical dews in the air. The symptoms include fever, dry cough, and weariness. Besides, aches and sore throat, and smallness of smell. Very rare persons have skilled diarrhea, biliousness or a liquid nose. Peoples having trouble in breathing or high fever cough need to pursue medical help immediately. The transmission of the disease among humans beings is exponentially cumulative the totals of the infested persons. The progress retro of this disease is one to fourteen days, or longer [5, 6]. As the expansion period of the virus is long it is not easy to analyze the optimal time required to monitor a curfew. Therefore, it is unsafe to revoke the curfew early [7, 8].

Artificial Neural Network (ANN) is a computing system planned to mimic the way that the human brain investigates and processes the incoming information. The method can solves problems that are not easy to tackle by human, or hard to analyze using standard statistical approaches. Training the ANN using past data permitting creation outputs according to the information mined from these data. ANN has a capability to generalize results from collection data that are unnoticed. Back propagation neural network (BPNN) is commonly used in several applications [7, 9]. In this method the algorithm used first for training, then for validating set. This set is utilized to check the trained network capability and perform a learning algorithm that makes a gradient descent optimization for making nodes weights the linking for every layer. BPNN is simple to use and robust, it is proved to offer good results in most cases.

Previous works benefiting from applying ANN in the medical field have been acknowledged by Pattichis and Pattichis, 2001 [10, 11]; Lei and Cheng, 2010; and Grossi and Buscema, [12-14]. Teng and Wah [1] accessible two learning devices for ANN that is beneficial to resolve binary classification problems, e.g., two-region, two-spiral, and Pima Indian Diabetes Diagnosis problems. The method is used for pattern classification into multiple categories by Xu and Chaudhari [15, 16]. Janghel, Shukla, Tiwari and Tiwari [17] developed neural network procedure using the pathological features for Clinical Decision Support System (CDSS) to evaluate the fatal delivery decision; normal or by surgical. Benjamin, Altman, O'Gorman, Rodeman and Peaz [18], use ANN for engineering analysis of elaborate physical systems. Simões, Furukawa, Mafra and Adamowski [19, 20]. ANN is used for acoustic signals classification as two classes of binary transmission. Bhatikar and Mahajan [21] also verified the use of the ANN for burden diagnosis of industrial apparatus, and how its learning benefits the assisting of an event with its process signature, at that point noticed and categorized events dependably. Soda, Pechenizkiy, Tortorella and Tsymbal [22] documented the structure of the precise features, feature removal, validating suitable methods of pattern illustration and dimensionality reduction is habitually more significant than the choice of the exact technique of classification.

## MATERIAL AND METHODS:

### Technical Background

Forecasting techniques might be instructed as a supplementary to the government in scheming enhanced plans and in creation creative choices. These methods measure the circumstances of the previous period, allowing good estimates for the state to happen in the future. Such forecasting will support governments worldwide to make for the approaching countries. In these forecasting methods a substantial role can be used for yielding closely accurate predictions [2].

### Regression Model [2]

Regression analysis can be used to study the association among dependent and independent variables. So, linear regression is a method for modeling the association between a dependent variable (or scalar reply) and unique or extra independent variables (or explanatory variables).

Consider a set of data of *n* numerical parts, it is assumed that the association between *y* (the dependent variable) and the *p*-vector of repressors x is linear, therefore a linear regression model is used [9]. And modeled by a commotion term (or error variable $\epsilon$ ), an unnoticed random variable that enhances "noise" to the linear association between the dependent variable and repressors. Therefore, the general form is expressed as:

$$yi = \beta_0 + \beta_1 x_{i1} + \cdots + \beta_p x_{ip} + \varepsilon_i = x_i^T \beta + \varepsilon_i \quad , i = 1, \ldots, n \quad (1)$$

Where T represents the transfer, so that $x_i^T \beta$ is the internal product between vectors xi and β.

While, polynomial regression that x (independent variable) , y (dependent variable) is modeled as in *n*th degree polynomial in x, a polynomial regression for is used for COVIE-19 predictions [23].

## Artificial Neural Network (ANN)

ANN mimics how the way biological nervous schemes procedure performed. The method including two phases: the learning and recall phases. In the learning phase, training signals (known data sets) are utilized to obtain updated weights in the input and output layers. The recall phase uses the weight gotten in the learning phase by one pass [24]. BPNN model includes; (1) The independent variables (or input layer nodes), (2) the dependent variable (or the output layer nodes), and (3) the hidden layer nodes. Back propagation (BP) is the learning or training algorithm rather than the network itself. Herein, batch mode training is performed in the back propagation training algorithm. The BP algorithm used in this approach is briefed as follows

1. All weights Initialization to small random values, usually ∈ [0,] using MATLAB.
2. Repeat the procedure until reaches termination criterion:
    2.1 A training set through the network is granted and forward pass across the network.
    2.2 Compute the actual output (o)
    2.3 Weights are modified; initially from the output layer, and then working backwards (backward pass)

The efficient weight between a node *p* and a node *q* at time t + 1 I, if node *p* is the *p*th node in a layer and node *q* is the *q*th node in the following layer in the forward way, ,

$$W_{pq}^{(t+1)} = W_{pq}^{(t)} + \Delta W_{pq} \quad (2)$$

where weight change, $\Delta W_{pq} = \eta . \delta_q . o_p$; η is the knowledge rate, δq is the error signal in the node q and $o_p$ is the yield of the node p. If neuron i is the i$^{th}$ node in the yield layer, the error sign in the output node i is,

$$\delta_i = (d_i - o_i) . o_i . (1 - o_i) \quad (3)$$

where $o_i$ and $d_i$ are the actual and desired output, respectively, of i$^{th}$ node in the yield layer, correspondingly If neuron j is the j$^{th}$ node in the layer which is hidden , the error signal in the layer which is hidden node j is,

$$\delta_j = o_j . (1 - o_j) . \sum (w_{ji} . \delta_i) \quad (4)$$

where $o_j$ is the yield of j$^{th}$ node in the layer which is hidden , $w_{ji}$ is the weight between j$^{th}$ node and i$^{th}$ node in the following layer, $\delta_i$ is the error sign in the i$^{th}$ node and the addition is over all the nodes in the following layer to the layer which is hidden .

In the trial, the ending standard is when the maximum number of epoch is touched. It classically receipts hundreds or thousands of epochs for an NN to converge [24]. In contrast, the ANN model can model compound scientific and logical systems with good fit results:

$$Y_t = F(X_{tn}) + \varepsilon_t \qquad (5)$$

where $Y_t = Y_1, Y_2, \ldots, Y_t$ is the response variable, $X_{tn}$ is a vector of *n* explanatory variables.

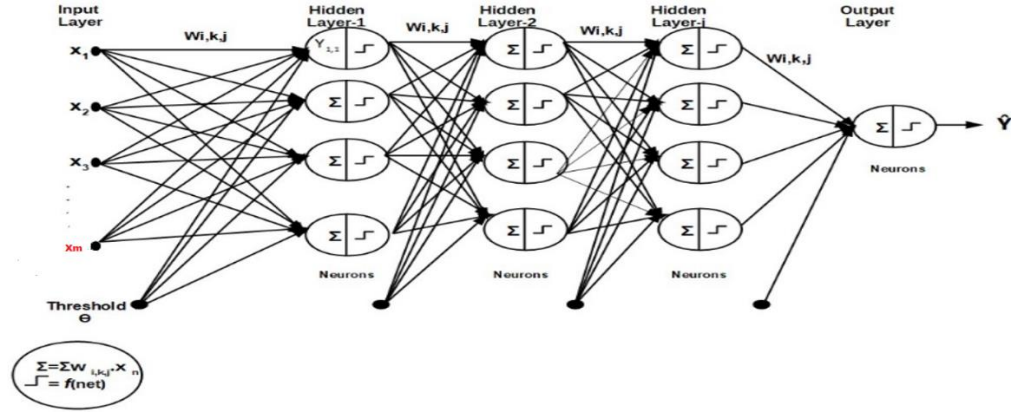

**Figure 1: Model of artificial neuron**

The backpropagation calculations involve three stages; including *Training*, *Validation/Test* and *Forecasting*. The dataset is split into three parts according the stated stages; the major part is used for training, the second part to validate the model accuracy, and the third for model testing. The steps of backpropagation are summarized through the *Training* stage as follows: for a certain learning rate ($\eta$), the weights are updated using a learning algorithm (gradient descent). The weights are adjusted anywhere in the network after the error becomes known. Then, the derived of the error equation with respect to that weight necessity be found and the delta $\Delta w$ is generated .
The nonlinear sigmoid function is used as a normalization function in the multi-layered perceptron models [25, 26]. The derivative being $\frac{dY}{dX} = \hat{Y} = \hat{Y}(1 - \hat{Y})$ and the delta rule using to change the weight $W_i$ in neuron k, hidden layer j with a sigmoid function.
Each training session involves attempting of different numbers of neurons and hidden layers for model improvement and outputs accuracy purposes. So, training session stops when the MSE is least. Therefore, the MLP is considered as the best model and set to be tested, by revealing it and compared to the test part of the data. After the training stage is completed, the model is now ready to validate then forecast future values.
Then, it used an adjustment factor to capture the difference between the behavior of actual datasets (used to train the network) and any deviated behavior caused by the new/future datasets in order to forecast. Any deviation from the network will be captured by an adjustment factor **Max** $(Y_a^{adj})$ or **Min** $(Y_a^{adj})$ that will be added to or subtracted from the forecasting formulation/model (3) based on the deviation type (positive or negative). A positive deviation $\Delta$**Max** $(Ya)$ means that the generated values are overestimated forecasts, while a negative deviation $\Delta$ **Min** $(Ya)$ means that the provided forecasts are underestimated. The adapted forecasting equations are [8, 25]:

$$\hat{Y}_{Forecast} = \left((\hat{Y_d})(Max\left(Y_a^{adj}\right) - Min\left(Y_a\right))\right) + Min\left(Y_a\right) \qquad (6)$$

- In case Min $(Y_a)$ is suggested for adjustment , the adapted forecasting equation is :

$$Y^{\hat{}}_{Forecast} = ((Y_d^{\hat{}})(Max(Y_a) - Min(Y_a^{adj}))) + Min(Y_a^{adj}) \qquad (7)$$

- In case both Min $(Y_a)$ and Min $(Y_a)$ are adjusted, the adapted forecasting equation is:

$$Y^{\hat{}}_{Forecast} = ((Y_d^{\hat{}})(Max(Y_a^{adj}) - Min(Y_a^{adj}))) + Min(Y_a^{adj}) \qquad (8)$$

$$Y^{\hat{}}_{Forecast} = ((Y_d^{\hat{}})(Max(Y_a^{adj}) - Min(Y_a))) + Min(Y_a)$$

## RESULTS

### Analysis and prediction for COVID-19 disease using a Sample of available Data

The data collected from Al Kindy teaching Hospital from 28/5/2019 to 28/7/2019 which focused on the daily figures of the main variables of our study interest: daily confirmed cases as a group, deaths of these cases, sex ( number of man and woman ), the number of patients under the age of 45 and over, and Number of patients with other diseases.

There are different techniques of forecasting has been used in literature based on purpose of forecasting and different data sources. It is shown in Fig. 2. That the deaths increased and decreased significantly with the confirmed cases.

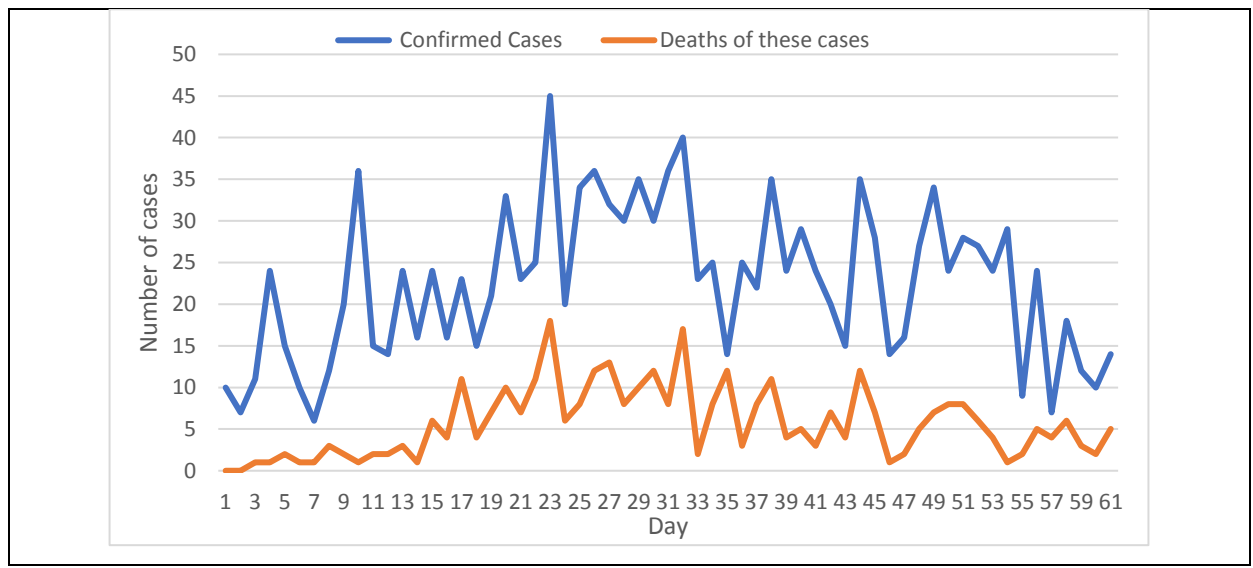

Fig 2. Number of daily covid 19 cases in terms of confirmed cases and mortality

### Model building and Forecasting

The data size used is categorized into three groups, 46 is selected as a training phase 46, 10 used for verification, 57 to 61 are used for comparison with the out-of-sample prediction test. The number of iterations used are 1000 for getting the optimal model.
Fig 3 The curve fitting of ANN model.

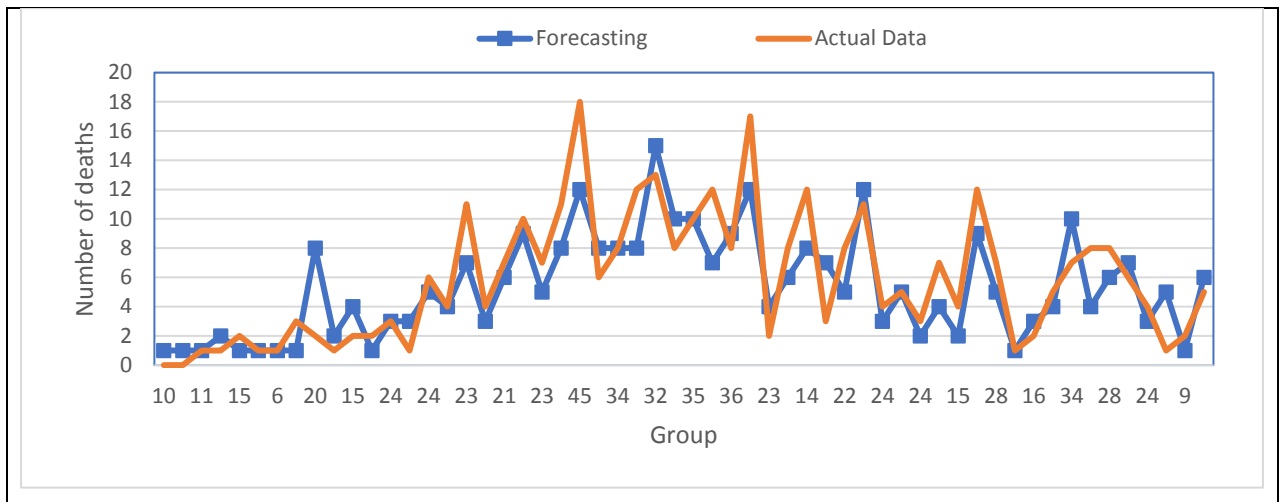
**Fig 3 : The curve fitting of ANN model.**

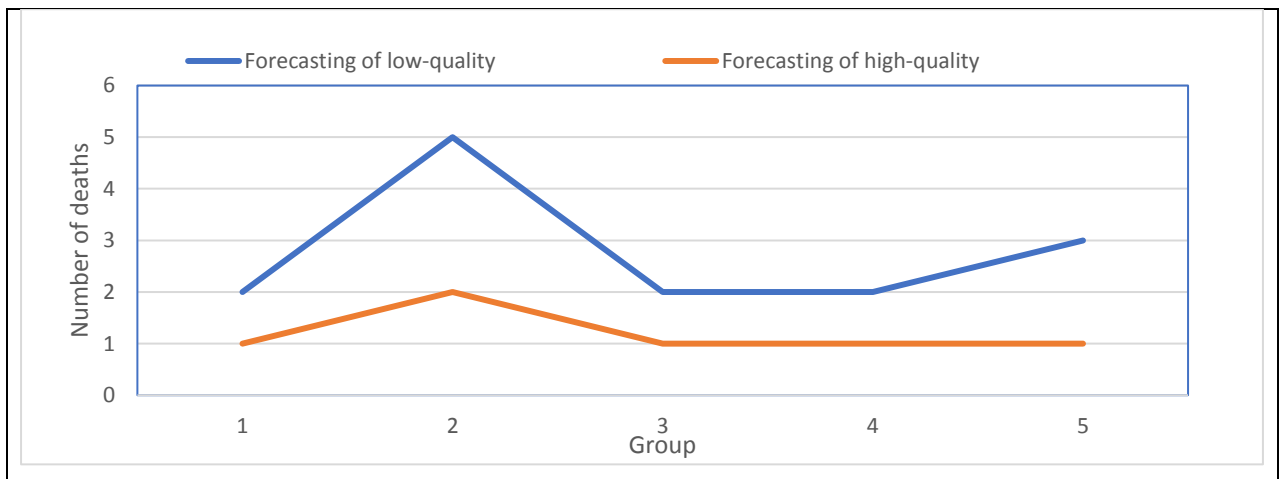
**Fig 4: comparison between high and low quality forecasting**

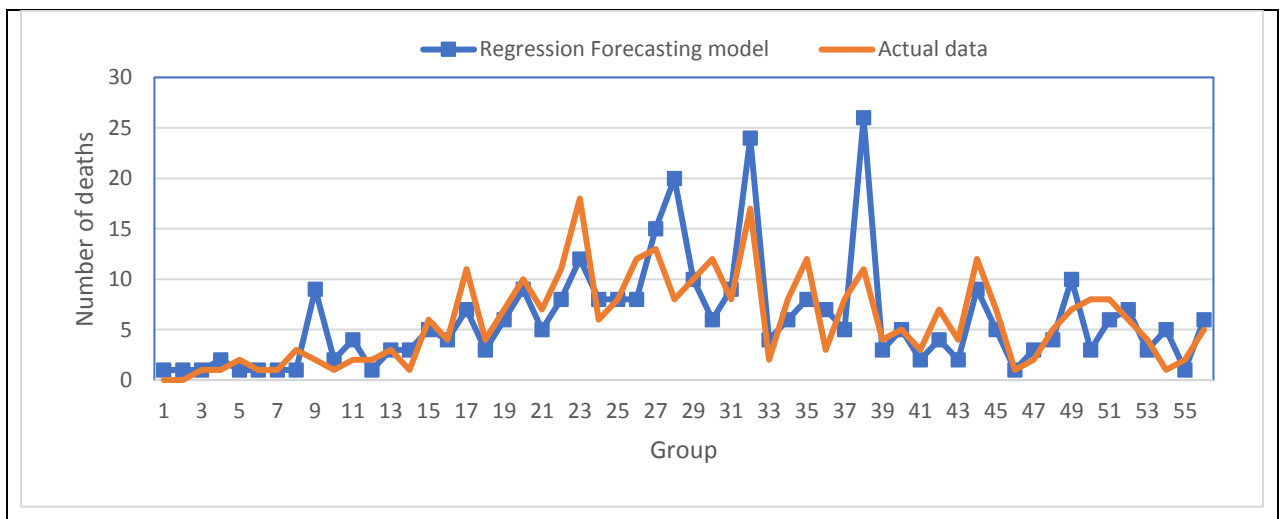
**Fig 5 : The multiple regression model used to obtain the fitting and validation forecasting model.**

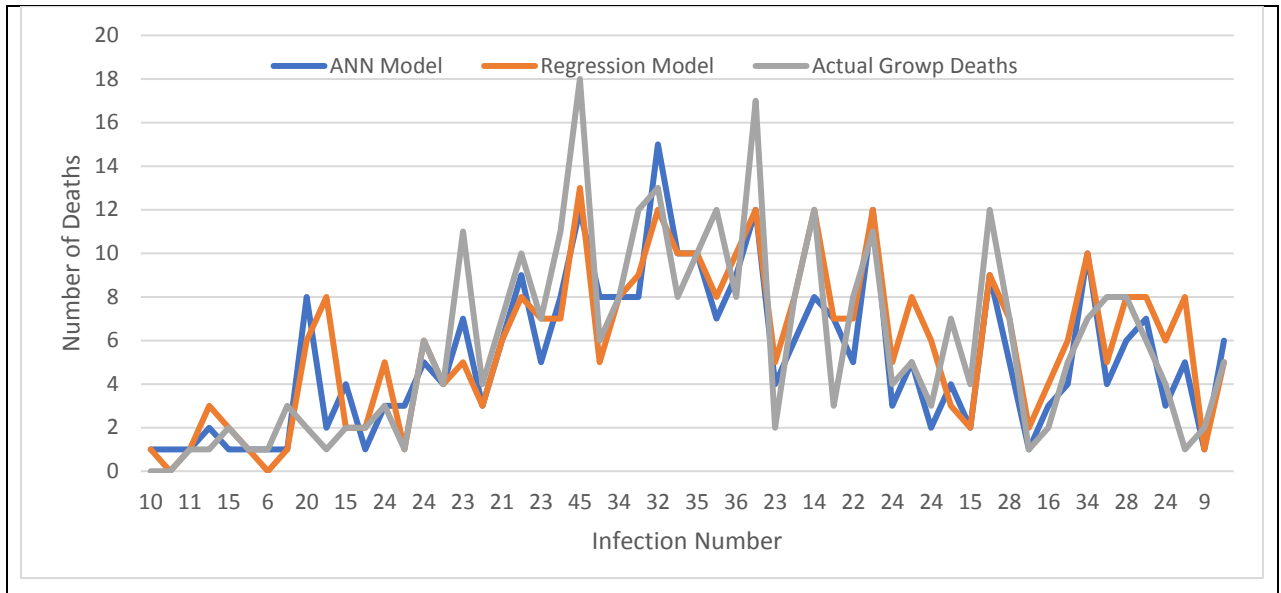

**Fig 6: comparison between ANN , regression and actual group deaths models.**

It is evident that the performance of the regression model in the case of dependence and exclusion status of a variable representing the case of chronic diseases is not sensitive.

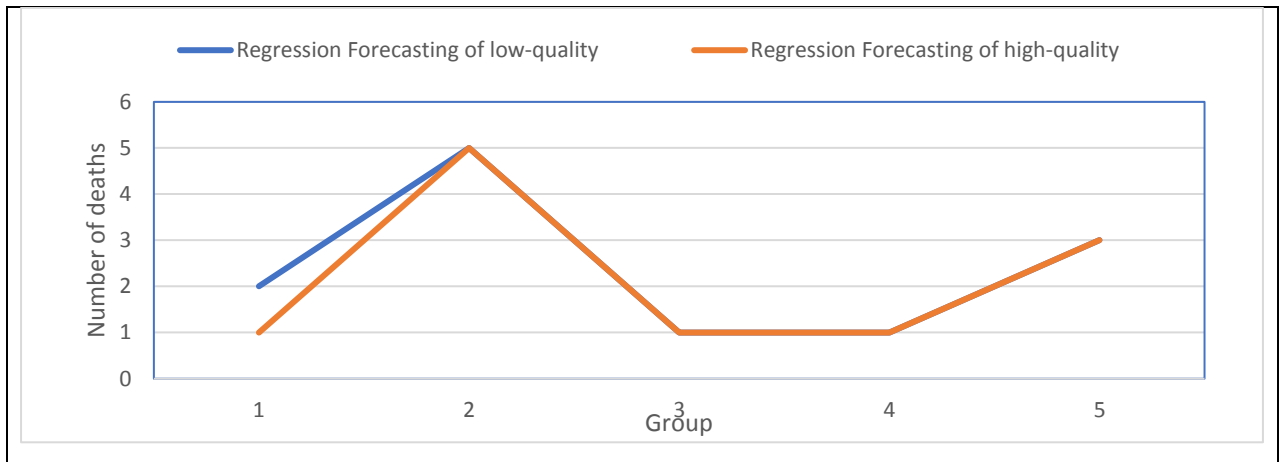

**Fig 7: comparison between high and low quality forecasting regression**

**Table 1 shows the linear regression (MAPE and MSE) and the ANN models in the training and validation stages.**

**Table 1. Comparison between the statistical indicators for the two models.**

|  | **Fitting** |  | **Validation** |  |
| --- | --- | --- | --- | --- |
| **Model** | MSE | MAPE | MSE | MAPE |
| **Regression** | 6.07 | 0.59 | 2.8 | 0.447 |
| **Neural network** | 5.91 | 0.46 | 4.40 | 0.52 |

It is clear in table 1 the poor performance of the traditional BP algorithm in long-term forecasting. Due to optimal weights obtained after the training stage has been completed might become less sustainable in long-term compared with short-term forecasting.

**Table 2. Comparison between the actual data with outputs for the two models**

| Number of deaths | Forecasting | |
|---|---|---|
| Actual data | ANN | Regression |
| 4 | 2 | 2 |
| 6 | 4 | 5 |
| 3 | 1 | 1 |
| 2 | 1 | 1 |
| 5 | 2 | 3 |

However, in order to improve the long-term forecasts further, the traditional forecast model/formulation used in the BPA (Equation 6-8) is improved by adding/subtracting the deviation represented by the adaptive factor to/from the forecast values obtained using Equation above. This adaptive factor is equal to the deviation between the behavior of long-term forecasts and the behavior theme the network becomes familiar with after the training stage have been completed as in table (3).

**Table 3**

| Actual data | Traditional ANN | Adjustment Applied |
|---|---|---|
| 4 | 2 | 2 |
| 6 | 4 | 5 |
| 3 | 1 | 2 |
| 2 | 1 | 2 |
| 5 | 2 | 3 |

With MAPE=.28 and MSE=2.0

**CONCLUSION:**

Accurately predicting the COVID-19 suitcases or death is crucial in determining making on the interference or regulator plan. This would expedite the selection of a suitable model used for forecasting. In this paper the ANN is used for the epidemic forecasting. The custom of ANN is not limited for prediction procedure, but it contains optimization, classification, estimation of the parameter , and best disease control organization. This research concludes the:

1- The deaths increased and decreased significantly with the confirmed cases.
2- It is evident that the performance of the regression model in the case of dependence and exclusion status of a variable representing the case of chronic diseases is not sensitive.
3- As shown in table (1) the poor performance of the traditional BP algorithm in long-term forecasting. Due to optimal weights obtained after the training stage has been completed might become less sustainable in long-term compared with short-term forecasting.

## References:

**1**. Teng C-C, Wah BW. Automated learning for reducing the configuration of a feedforward neural network. IEEE transactions on neural networks. 1996;7(5):1072-85
**2**. Gola A, Arya RK, Dugh R. Review of Forecasting Models for Coronavirus (COVID-19) Pandemic in India during Country-wise Lockdown. MedRxiv. 2020
**3**. Mor S, Saini P, Wangnoo SK, Bawa T. Worldwide spread of COVID-19 Pandemic and risk factors among Co-morbid conditions especially Diabetes Mellitus in India. Research Journal of Pharmacy and Technology. 2020;13(5):2530-2
**4**. Patil PA, Jain RS. Theoretical Study and treatment of Novel COVID-19. Research journal of Pharmacology and Pharmacodynamics. 2020;12(2):71-2
**5**. Sujath R, Chatterjee JM, Hassanien AE. A machine learning forecasting model for COVID-19 pandemic in India. Stochastic Environmental Research and Risk Assessment. 2020;34:959-72
**6**. Ahmad S, Shoaib A, Ali S, Alam S, Alam N, Ali M, et al. Epidemiology, risk, myths, pharmacotherapeutic management and socio economic burden due to novel COVID-19: A recent update. Research Journal of Pharmacy and Technology. 2020;13(9):4435-42
**7**. Philemon MD, Ismail Z, Dare J. A review of epidemic forecasting using artificial neural networks. International Journal of Epidemiologic Research. 2019;6(3):132-43
**8**. Ingole RD, Thalkari AB, Karwa PN. Is Prevaccination the reason for less morbidity and mortality for COVID-19 in India: An Epidemiological study. Research Journal of Science and Technology. 2020;12(4):285-8
**9**. Dogan SZ, Arditi D, Gunaydin HM. Comparison of ANN and CBR models for early cost prediction of structural systems: Bauhaus-Universität; 2006.
**10**. Pattichis CS, Pattichis MS, editors. Adaptive neural network imaging in medical systems. Conference Record of Thirty-Fifth Asilomar Conference on Signals, Systems and Computers (CatNo01CH37256); 2001 4-7 Nov. 2001.
**11**. Dewangan V, Sahu R, Satapathy T, Roy A. The exploring of current development status and the unusual symptoms of coronavirus pandemic (Covid-19). Research journal of Pharmacology and Pharmacodynamics. 2020;12(4):172-6
**12**. Grossi E, editor Artificial Adaptive Systems and predictive medicine: a revolutionary paradigm shift. Immunity & Ageing; 2010: Springer.
**13**. Kishor RS, Ramhari BM. Introduction to Covid-19. Research Journal of Science and Technology. 2020;12(4):338-45
**14**. Shankhdhar PK, Mishra P, Kannojia P, Joshi H. Turmeric: Plant immunobooster against covid-19. Research Journal of Pharmacognosy and Phytochemistry. 2020;12(3):174-7

**15.** Xu Y, Chaudhari NS. Application of binary neural networks for classification. Proceedings of the 2003 International Conference on Machine Learning and Cybernetics (IEEE Cat No03EX693). 2003;3:1343-8 Vol.3

**16.** Kumar R, Chawla A. A valuable insight to the novel deadly covid-19: A review. Research journal of Pharmacology and Pharmacodynamics. 2020;12(3):111-6

**17.** Janghel RR, Shukla A, Tiwari R, Tiwari P. International conference on new trends in information and service science.: IEEE Computer Society; 2009. p. 170-5.

**18.** Benjamin A, Altman B, o Gorman C, Rodeman R, Paez TL. Use of artificial neural networks for engineering analysis of complex physical systems. Proceedings of the Thirtieth Hawaii International Conference on System Sciences. 1997;5:511-20 vol.5

**19.** Simoes MG, Furukawa CM, Mafra AT, Adamowski JC. A novel competitive learning neural network based acoustic transmission system for oil-well monitoring. IEEE Transactions on Industry Applications. 2000;36(2):484-91.10.1109/28.833765

**20.** Derouiche S. Current Review on Herbal Pharmaceutical improve immune responses against COVID-19 infection. Research Journal of Pharmaceutical Dosage Forms and Technology. 2020;12(3):181-4

**21.** Bhatikar SR, Mahajan RL. Artificial neural-network-based diagnosis of CVD barrel reactor. IEEE transactions on semiconductor manufacturing. 2002;15(1):71-8

**22.** Soda P, Pechenizkiy M, Tortorella F, Tsymbal A. Guest editorial: Knowledge discovery and computer-based decision support in biomedicine. Artificial intelligence in medicine. 2010;50(1):1-2

**23.** Shi L, Wang X-c. Artificial neural networks: Current applications in modern medicine2010. 383-7 p.10.1109/CCTAE.2010.5543470

**24.** Biswas S, Baruah B, Purkayastha B, Chakraborty M. An ANN based Classification Algorithm for Swine Flu Diagnosis. International Journal of Knowledge Based Computer Systems. 2016;3.10.21863/ijkbcs/2015.3.1.005

**25.** Picton P. Neural Networks: Palgrave Macmillan; 2001.

**26.** Kuser AK, Tarar SM, Kassid OM, Zayer NH. CNS and COVID-19: Neurological symptoms of Hospitalized Patients with Coronavirus in Iraq: a surveying case sequences study. Research Journal of Pharmacy and Technology. 2020;13(12):6291-4